# Proton-Electron Mass Ratio from Laser Spectroscopy of HD+ at the Part-Per-Trillion Level

Sayan Patra[1], M. Germann[1], J.-Ph. Karr[2,3], M. Haidar[2], L. Hilico[2,3], V. I. Korobov[4], F. M. J. Cozijn[1], K. S. E. Eikema[1], W. Ubachs[1], J. C. J. Koelemeij[1]*

[1] LaserLaB, Department of Physics and Astronomy, Vrije Universiteit Amsterdam, De Boelelaan 1081, 1081 HV Amsterdam, The Netherlands.

[2] Laboratoire Kastler Brossel, UPMC-Sorbonne Université, CNRS, ENS-PSL Research University, Collège de France, 4 place Jussieu, 75005 Paris, France.

[3] Université d'Evry-Val d'Essonne, Université Paris-Saclay, Boulevard François Mitterrand, 91000 Evry, France.

[4] Bogolyubov Laboratory of Theoretical Physics, Joint Institute for Nuclear Research, Dubna 141980, Russia.

*Correspondence to: j.c.j.koelemeij@vu.nl

**Accepted values of the masses of several subatomic particles have been under debate since recent measurements in Penning traps produced more precise yet incompatible results, implying possible inconsistencies in closely related physical constants like the proton-electron and deuteron-proton mass ratios. These quantities also influence the predicted vibrational spectrum of the deuterated molecular hydrogen ion in its electronic ground state, of which we measured the $v=0 \rightarrow 9$ overtone transition frequency with an uncertainty of 2.9 parts-per-trillion through Doppler-free two-photon laser spectroscopy. Leveraging high-precision *ab initio* calculations we convert our measurement to tight constraints on the proton-electron and deuteron-proton mass ratios, consistent with the most recent Penning-trap determinations of these quantities, and yielding a new value of the proton-electron mass ratio with an unprecedented precision of 21 parts-per-trillion.**

Precision measurements on subatomic particles and simple atomic systems play an essential role in the determination of fundamental physical constants. Examples range from the proton-electron mass ratio, $m_p/m_e$, whose value depends strongly on measurements performed on single protons and electrons stored in Penning traps, to the Rydberg constant, $R_\infty$, and proton electric charge radius, $r_p$, which are derived from spectroscopic measurements of energy intervals in atomic hydrogen-like systems (*1,2*). It is desirable to perform such determinations of physical constants redundantly using different systems and methods, as this provides a crucial cross-check for possible experimental inconsistencies or shortcomings in our understanding of nature. This is illustrated by the "Proton Radius Puzzle," a 5.6σ discrepancy between the value of $r_p$ obtained from muonic-hydrogen spectroscopy and the 2014 Committee on Data for Science and Technology (CODATA-2014) reference value (*1,3*). While this puzzle now seems solved with most recent $r_p$ determinations from electron-proton scattering and atomic-hydrogen spectroscopy converging to the muonic-hydrogen value (*4-7*), other inconsistencies continue to exist. These concern the (atomic) masses of the proton, $m_p$, deuteron, $m_d$, and helion, $m_h$, which have been determined through Penning-trap mass measurements, but with recent results revealing discrepancies of several standard uncertainties with earlier values (*8-15*). For example, Heiße *et al.* determined $m_p$ with 32 parts-per-trillion (ppt) precision, three times higher than the then-accepted CODATA-2014 value, but also found it to be smaller by 3σ (*11,12*). The value from (*11*) has been incorporated in the 2017 and forthcoming 2018 CODATA adjustments, but uncertainty margins were increased by a factor of 1.7 to accommodate the discrepancy (*2*). This currently limits the precision of $m_p/m_e$ (obtained by dividing $m_p$ by the more precise CODATA-2018 value of $m_e$) to 60 ppt. This in turn diminishes the predictive power of *ab initio* calculations of rotational-vibrational spectra of molecular hydrogen ions ($H_2^+$, $HD^+$) and antiprotonic helium, which have achieved a precision of 7-8 ppt (*16*).

By the same token, this situation opens up the prospect of determining a new and improved value of $m_p/m_e$ from measurements of these systems, which could shed light on the proton-mass discrepancy (*17*). However, this requires measurements with ppt-level uncertainties, two orders of magnitude beyond state-of-the-art laser (*18,19*) and terahertz (*20*) spectroscopy of HD⁺ and antiprotonic helium. Here, we present a frequency measurement of the (*v,L*): (0,3) → (9,3) vibrational transition in the electronic ground state of HD⁺ with 2.9 ppt uncertainty, allowing us to derive a new value of $m_p/m_e$ with unprecedented precision, while providing a cross-link to other physical constants which enables additional consistency checks of their values.

We previously identified the (*v,L*): (0,3) → (4,2) → (9,3) two-photon transition in HD⁺ (Fig. 1A) as a promising candidate for high-resolution Doppler-free laser spectroscopy (*21*), owing to the near degeneracy of the 1442 nm and 1445 nm photons involved, and the possibility to store HD⁺ ions in a linear Paul trap while cooling them to 10 mK through Coulomb interaction with co-trapped beryllium ions, themselves cooled by 313 nm laser radiation. We showed that for counter-propagating 1442 nm and 1445 nm laser beams directed along the trap's symmetry axis, Doppler-free vibrational excitation of HD⁺ deep in the optical Lamb-Dicke regime may be achieved. Thus, with a natural linewidth of 13 Hz, $Q$-factors of >$10^{13}$ come within reach. We use phase-stabilized, continuous-wave external cavity diode lasers at 1442 nm and 1445 nm possessing line widths of 1−2 kHz to vibrationally excite cold, trapped HD⁺ ions (*22*). Optical frequencies are measured with an uncertainty below 1 ppt using an optical frequency comb laser, while two-photon excitation is detected through enhanced loss of HD⁺ from the trap due to state-selective dissociation of molecules in the *v*=9 state by 532 nm laser radiation (*22,23*).

Rovibrational energy levels of HD⁺ exhibit hyperfine structure due to magnetic interactions between the spins of the proton, $\mathbf{I}_p$, deuteron, $\mathbf{I}_d$, and electron, $\mathbf{s}_e$, as well as the molecule's rotational angular momentum, $\mathbf{L}$ (*24*). The spins are coupled to form resultant angular momenta $\mathbf{F}=\mathbf{s}_e+\mathbf{I}_p$ and $\mathbf{S}=\mathbf{F}+\mathbf{I}_d$, and are finally coupled with $\mathbf{L}$ to form the total angular momentum $\mathbf{J}=\mathbf{S}+\mathbf{L}$. Here, we observe transitions $(v,L; F,S,J)$: $(0,3;1,2,5) \rightarrow (9,3;1,2,5)$ (henceforth referred to as the '$F=1$ transition'), and $(v,L; F,S,J)$: $(0,3;0,1,4) \rightarrow (9,3;0,1,4)$ (referred to as the '$F=0$ transition'); see Fig. 1B.

To record a spectrum, we keep the 1442 nm laser frequency, $\nu_F$ (with $F=0,1$; see Fig. 1B), at a fixed detuning $\delta_F$ from resonance to avoid excessive population of the intermediate $v=4$ state (*21,22*). Meanwhile, we step the 1445 nm laser frequency, $\nu'_F$, in intervals of 2 kHz over the range of interest (Fig. 1B). At each step we let all lasers interact with the $HD^+$ ions for 30 s, after which we determine the cumulative loss of $HD^+$, and add the resulting data point to the spectrum (*22*). A typical spectrum covers a span of 40 to 60 kHz with on average nine points per frequency, and with the 180 to 270 data points acquired in random order over the course of about ten measurement days. The signal-to-noise ratio of the $F=0$ spectrum turned out to be lower than its $F=1$ counterpart, which we attribute to smaller available population in the initial state, and slower re-population by blackbody radiation (*21*). To increase the $F=0$ signal, we apply two radio-frequency (rf) magnetic fields driving population from the $(F,S,J)=(1,2,5)$ and $(1,2,4)$ states of the $v=0$, $L=3$ hyperfine manifold to the $(F,S,J)=(0,1,4)$ states; see Figs. 1B and S1 (*22*). Recorded spectra of the $F=0$ and $F=1$ transitions are shown in Fig. 2.

The interpretation of the recorded spectra requires an analysis of several systematic effects which affect the line shape and position (*22*). Here we exploit the good theoretical accessibility of the $HD^+$ molecule (*25*), which allows *a priori* estimation of these effects. Zeeman and Stark effects are calculated to shift the $F=0$ and $F=1$ lines by up to 0.5 kHz, which occurs through level shifting as well as line shape deformation (*22*). Expected two-photon power broadening and interaction-time broadening due to the $9 \times 10^3$/s rate of dissociation of molecules in the $v$=9 state (*21*) satisfactorily explain the observed line widths of 8(3) kHz. In addition, we experimentally investigated a number of systematic effects, yielding results consistent with the theory-based estimates (*22*). The size and uncertainty of leading systematic effects are listed in Table 1.

As shown in Fig. 2, Lorentzian line shapes are fitted to the spectra to find their respective line centers with 0.6–0.7 kHz uncertainty. These are subsequently corrected for systematic frequency shifts, and combined to arrive at the $F$=0 and $F$=1 transition frequencies, $\nu_{HF0,exp}$ and $\nu_{HF1,exp}$ (*22*); see Figs. 2C,D and Table 2. These frequencies are related to the spin-averaged (i.e. pure rovibrational) frequency, $\nu_{SA}$, through the relations $\nu_{SA} = \nu_{HF0} - f_{0c}$ and $\nu_{SA} = \nu_{HF1} - f_{1c}$ (Fig. 1C). Since only $\nu_{SA}$ depends directly on the values of the physical constants of interest, we need to determine and correct for the hyperfine shifts, $f_{1c} \approx -63$ MHz and $f_{0c} \approx 115$ MHz to derive $\nu_{SA}$. We take the hyperfine intervals, $f_{0c,theo}$ and $f_{1c,theo}$ from theory (*22,24,26*), and compute $\nu_{SA,exp}$ as the mean of $\nu_{HF0,exp} - f_{0c,theo}$ and $\nu_{HF1,exp} - f_{1c,theo}$ (*22*). In this process, we expand the uncertainties of the theoretical hyperfine intervals by about a factor of two (*22*), so that the theoretical hyperfine interval, $f_{10,theo}$, becomes consistent with its measured counterpart, $f_{10,exp} \equiv \nu_{HF0,exp} - \nu_{HF1,exp}$ (Table 2). We thus find $\nu_{SA,exp} = 415,264,925,500.5(0.4)_{exp}(1.1)_{theo}(1.2)_{total}$ kHz.

Our experimental frequency $\nu_{SA,exp}$ exceeds the theoretical frequency $\nu_{SA,theo}$ (CODATA-2014) = 415,264,925,467.1(10.2) kHz by 33.4 kHz, or 3.3σ, when we use CODATA-2014 physical constants to compute $\nu_{SA,theo}$ (*22,27*). The uncertainties of these constants dominate the 10.2 kHz uncertainty rather than the 3.1 kHz precision of the theoretical model; $m_p/m_e$ for example contributes 9.0 kHz (Fig. S3) (*22*). Using known sensitivity coefficients (*17,22*), we can also compute other theoretical frequency values, $\nu_{SA,theo}(k)$, for other combinations (labeled $k$) of values of physical constants. For example, a more precise value is obtained by use of CODATA-2018 constants, $\nu_{SA,theo}$ (CODATA-2018) = 415,264,925,496.2(7.4) kHz. This state-of-the-art value is shifted by 29.1 kHz with respect to the CODATA-2014 value (Fig. 3A), and essentially closes the 33.4 kHz gap with our experimental value $\nu_{SA,exp}$. Figure 3A furthermore shows that most of the 29.1 kHz shift stems from the smaller CODATA-2018 value of $m_p/m_e$. A smaller part, 5.1 kHz, is due to the CODATA-2018 updated values of $r_p$, $r_d$, and $R_\infty$, which are essentially equal to the muonic-hydrogen values (*3,28*). The 5.1 kHz shift, which is four times larger than our experimental uncertainty and comparable to the current theoretical precision, therefore reveals the impact of the Proton Radius Puzzle for the first time on molecular vibrations. We obtain even better precision (5.5 kHz) and agreement after replacing the CODATA-2018 value of $m_p/m_e$ with that from (*11,12*), this time leading to a 31.2 kHz shift (Fig. 3A).

We may also invert the procedure and derive a new value of $m_p/m_e$ from the difference $\nu_{SA,exp}-\nu_{SA,theo}$ (*k*); see Fig. 3B. Using $\nu_{SA,theo}$ (CODATA-2018), we obtain $m_p/m_e$ (HD$^+$) = 1,836.152 673 349(71) which is slightly more precise than, and in excellent agreement with, the value of $m_p/m_e$ from (*12*). With $\nu_{SA,theo}$ being also sensitive to the deuteron-proton mass ratio (*22*), one may alternatively extract a two-dimensional constraint in the ($m_p/m_e$, $m_d/m_p$) plane (Fig. 3C). Our result is found to be in good agreement with both $m_p/m_e$ from (*12*) and the recent value of $m_d/m_p$ (*14*), assuming CODATA-2018 values of $r_p$, $r_d$, and $R_\infty$. This justifies a determination of $m_p/m_e$ from all three results shown in Fig. 3C combined, leading to a value of 1,836.152 673 406(38) (bottommost point in Fig. 3B) which, at 21 ppt precision, represents the most precise determination of this quantity to date. The data shown in Fig. 3C can furthermore be combined with the CODATA-2018 value of $m_e$ and the value of $m_h$ from (*15*) to obtain the atomic mass difference $m_p + m_d - m_h$ = 0.005 897 432 54(12) u. The same quantity has previously been determined from the measured mass ratio $^3$He$^+$/HD$^+$ (*13*), leading to $m_p + m_d -$ $m_h$ = 0.005 897 432 19(7) u. The two results differ by 0.35(14) nu, or 2.5σ. We thereby confirm the "$^3$He puzzle," a term used to describe similar discrepancies of 0.48(10) nu (or 4.8σ) and 0.33(13) nu (or 2.4σ) reported earlier in (*13*) and (*14*), respectively.

Our work establishes precision spectroscopy of HD$^+$, combined with *ab initio* quantum-molecular calculations, as a state-of-the-art method for determining fundamental mass ratios. It furthermore provides a linking pin between mass ratios and other physical constants, such as $R_\infty$, and sheds new light on recently observed discrepancies. We therefore anticipate that our results will have a significant impact on the consistency and precision of future reference values of physical constants, and enhance the predictive power of *ab initio* calculations of physical quantities.

**References and Notes:**

1. P. J. Mohr, D. B. Newell, B. N. Taylor, CODATA Recommended Values of the Fundamental Physical Constants: 2014. *J. Phys. Chem. Ref. Data* **45**, 043102 (2016).

2. P. J. Mohr, D. B. Newell, B. N. Taylor, E. Tiesinga, Data and analysis for the CODATA 2017 special fundamental constants adjustment. *Metrologia* **55**, 125 -146 (2018).

3. A. Antognini *et al.,* Proton structure from the measurement of 2S-2P transition frequencies of muonic hydrogen. *Science* **339,** 417-420 (2013).

4. A. Beyer, L. Maisenbacher, A. Matveev, R. Pohl, K. Khabarova, A. Grinin, T. Lamour, D. C. Yost, T. W. Hänsch, N. Kolachevsky, T. Udem, The Rydberg constant and proton size from atomic hydrogen. *Science* **358**, 79-85 (2017).

5.  H. Fleurbaey, S. Galtier, S. Thomas, M. Bonnaud, L. Julien, F. Biraben, F. Nez, M. Abgrall, J. Guéna, New measurement of the $1S-3S$ transition frequency of hydrogen: contribution to the proton charge radius puzzle. *Phys. Rev. Lett.* **120,** 183001(2018).

6.  N. Bezginov, T. Valdez, M. Horbatsch, A. Marsman, A. C. Vutha, E. A. Hessels, A measurement of the atomic hydrogen Lamb shift and the proton charge radius. *Science* **365,** 1007-1012 (2019).

7.  W. Xiong *et al.*, A small proton charge radius from an electron–proton scattering experiment. *Nature* **575,** 147-150 (2019).

8.  R. S. Van Dyck, Jr., D. L. Farnham, S. L. Zafonte, P. B. Schwinberg, High precision Penning trap mass spectroscopy and a new measurement of the proton's "atomic mass." *AIP Conf. Proc.* **457,** 101 (1999).

9.  I. Bergström, T. Fritioff, R. Schuch, J. Schönfelder, On the masses of $^{28}$Si and the proton determined in a Penning trap. *Phys. Scr.* **66,** 201 (2002).

10. A. Solders, I. Bergström, S. Nagy, M. Suhonen, R. Schuch, Determination of the proton mass from a measurement of the cyclotron frequencies of $D^+$ and $H_2^+$ in a Penning trap. *Phys. Rev. A* **78,** 012514 (2008).

11. F. Heiße, F. Köhler-Langes, S. Rau, J. Hou, S. Junck, A. Kracke, A. Mooser, W. Quint, S. Ulmer, G. Werth, K. Blaum, S. Sturm, High-precision measurement of the proton's atomic mass. *Phys. Rev. Lett.* **119**, 033001 (2017).

12. F. Heiße, S. Rau, F. Köhler-Langes, W. Quint, G. Werth, S. Sturm, K. Blaum, High-precision mass spectrometer for light ions. *Phys. Rev. A* **100**, 022518 (2019).

13. S. Hamzeloui, J. A. Smith, D. J. Fink, E. G. Myers, Precision mass ratio of $^3He^+$ to $HD^+$. *Phys. Rev. A* **96,** 060501(R) (2017).

14. D. J. Fink, E. G. Myers, Deuteron-to-proton mass ratio from the cyclotron frequency ratio of $H_2^+$ to $D^+$ with $H_2^+$ in a resolved vibrational state. *Phys. Rev. Lett.* **124,** 013001 (2020).

15. S. L. Zafonte, R. S. Van Dyck, Jr., Ultra-precise single-ion atomic mass measurements on deuterium and helium-3. *Metrologia* **52,** 280 (2015).

16. V.I. Korobov, L. Hilico, and J.-Ph. Karr, Fundamental transitions and ionization energies of the hydrogen molecular ions with few ppt uncertainty. *Phys. Rev. Lett.* **118,** 233001 (2017).

17. J.-Ph. Karr, L. Hilico, J. C. J. Koelemeij, V. I. Korobov, Hydrogen molecular ions for improved determination of fundamental constants. *Phys. Rev. A* **94**, 050501(R) (2016).

18. J. Biesheuvel, J.-Ph. Karr, L. Hilico, K. S. E. Eikema, W. Ubachs, J. C. J. Koelemeij, Probing QED and fundamental constants through laser spectroscopy of vibrational transitions in $HD^+$. *Nat. Commun.* **7**, 10385 (2016).

19. M. Hori *et al.*, Buffer-gas cooling of antiprotonic helium to 1.5 to 1.7 K, and antiproton-to-electron mass ratio. *Science* **354**, 610-614 (2016).

20. S. Alighanbari, M. Hansen, V. I. Korobov, S. Schiller, Rotational spectroscopy of cold and trapped molecular ions in the Lamb-Dicke regime. *Nat. Physics* **14**, 555-559, (2018).

21. V.Q. Tran, J.-Ph. Karr, A. Douillet, J.C.J. Koelemeij, L. Hilico, Two-photon spectroscopy of trapped $HD^+$ ions in the Lamb-Dicke regime. *Phys. Rev. A* **88**, 033421 (2013).

22. Materials and methods are available as supplementary materials at the Science website.

23. J. Biesheuvel, J.-Ph. Karr, L. Hilico, K. S. E. Eikema, W. Ubachs, J. C. J. Koelemeij, High-precision spectroscopy of the $HD^+$ molecule at the 1-p.p.b. level *Appl. Phys. B* **123,** 1-22 (2017).

24. D. Bakalov, V. I. Korobov, S. Schiller, High-precision calculation of the hyperfine structure of the $HD^+$ ion. *Phys. Rev. Lett.* **97**, 243001 (2006).

25. J.-Ph. Karr, $H_2^+$ and $HD^+$: Candidates for a molecular clock. *J. Mol. Spec.* **300**, 37-43 (2014).

26. V. I. Korobov, J. C. J. Koelemeij, L. Hilico, J.-Ph. Karr, Theoretical hyperfine structure of the molecular hydrogen ion at the 1 ppm level. *Phys. Rev. Lett.***116**, 053003 (2016).

27. D. T. Aznabayev, A. K. Bekbaev, and V.I. Korobov, Leading-order relativistic corrections to the rovibrational spectrum of $H_2^+$ and $HD^+$ molecular ions. *Phys. Rev. A* **99,** 012501 (2019).

28. R. Pohl *et al.*, Laser spectroscopy of muonic deuterium. *Science* **353,** 669-673 (2016).

**Acknowledgments:** We are indebted to Rob Kortekaas, Tjeerd Pinkert, and the Electronic Engineering Group of the Faculty of Science at Vrije Universiteit Amsterdam for technical assistance.

**Funding:** We acknowledge support from the Netherlands Organisation for Scientific Research (FOM Programs "Broken Mirrors & Drifting Constants" and "The Mysterious Size of the Proton"; FOM 13PR3109, STW Vidi 12346), the European Research Council (AdG 670168 Ubachs, AdG 695677 Eikema), the COST Action CA17113 TIPICQA, and the Dutch-French bilateral Van Gogh program. J.-Ph. K. acknowledges support as a fellow of the Institut Universitaire de France. VIK acknowledges support of the Russian Foundation for Basic Research under Grant No.~19-02-00058-a.

**Supplementary Materials:**

Materials and Methods

Figures S1-S3

Tables S1-S3

References (*29-46*)

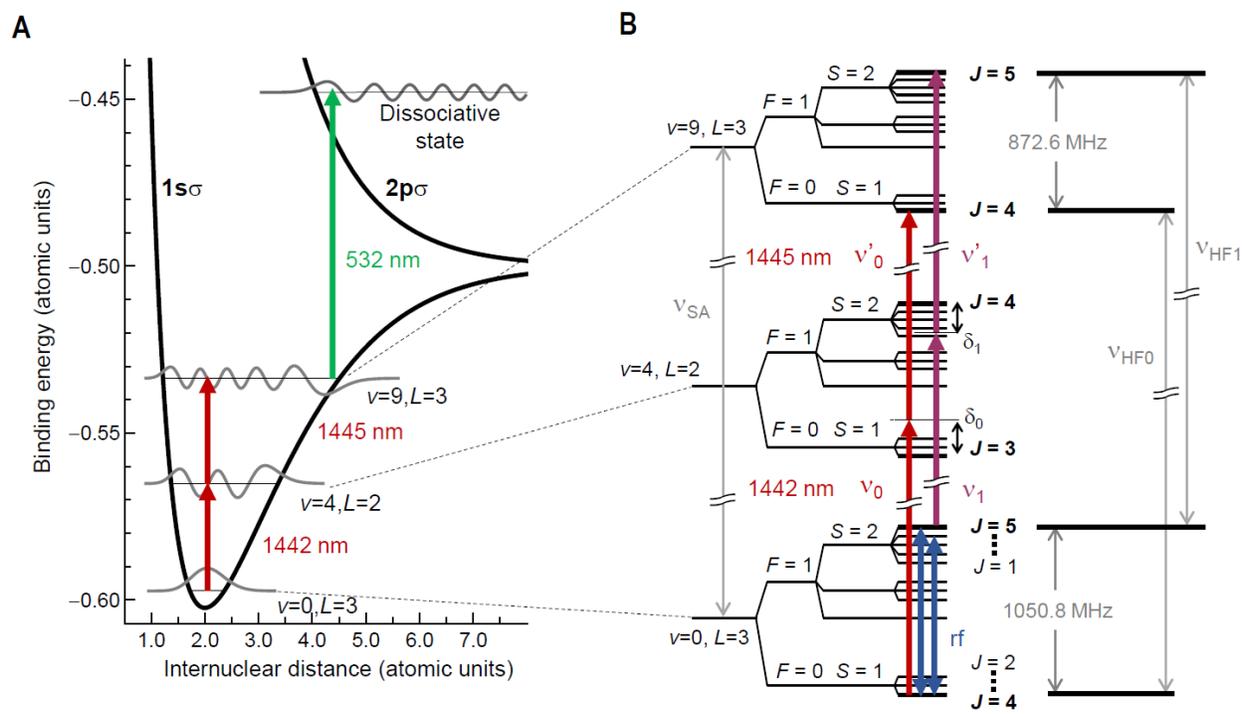

**Fig. 1. Partial level diagram and multi-photon transitions.** (**A**) Two-photon transitions are driven between rovibrational states with ($v$,$L$) = (0,3) and (9,3) in the 1sσ electronic ground state of HD⁺. State-selective dissociation of $v$=9 population is induced through excitation to the antibonding 2pσ electronic state by a 532 nm photon. (**B**) Spin-averaged transition frequency, $v_{SA}$, and hyperfine structure (not to scale) of the levels involved in the two-photon transition, and graphical definitions of the frequencies and detunings of the electromagnetic fields driving transitions between them. (**C**) Graphical definition of the hyperfine intervals in the two-photon transition.

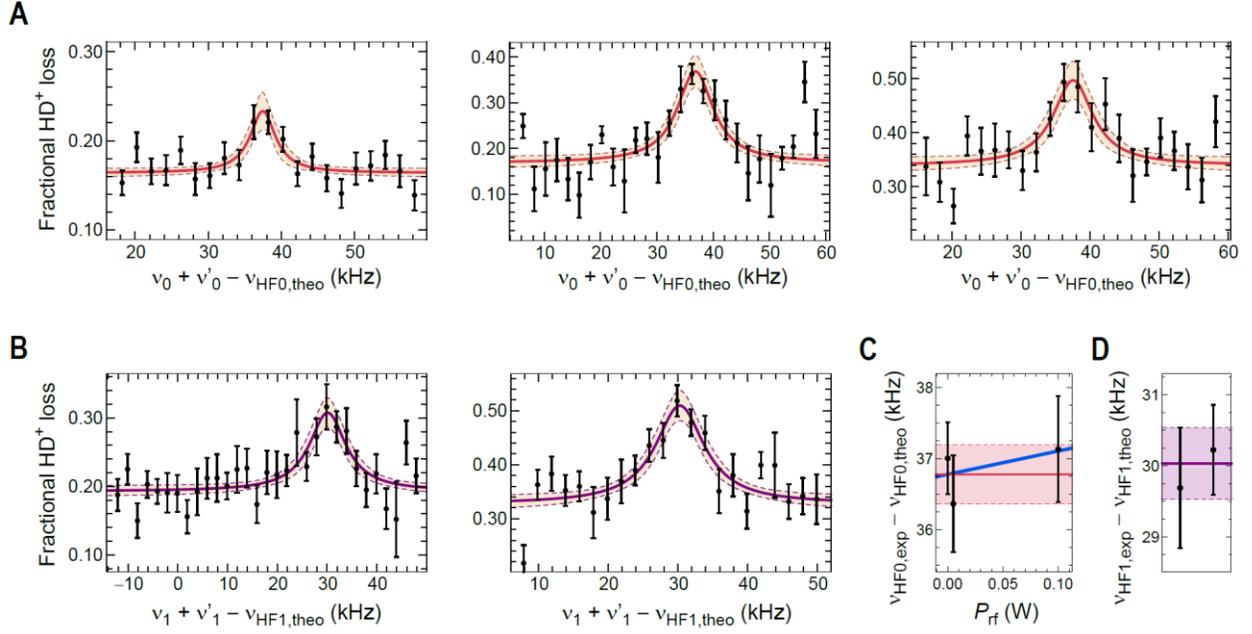

**Fig. 2. Spectra of the two-photon transition at 415 THz.** (**A**) Spectra of the $F$=0 transition at various rf power levels (from left to right): 0 mW, 5 mW, 100 mW. Lorentzian line fits are shown along with 68% confidence-level bands. Each data point represents the mean of a set of typically nine individual measurements, with the error bar indicating the standard error of the mean. (**B**) Spectral data and Lorentzian line fits for the $F$=1 transitions at two 532-nm-laser intensities: 2.5 MW/m$^2$ (left) and 0.57 MW/m$^2$ (right). (**C**) Fitted line centers of the $F$=0 transitions (corrected for systematic shifts (*22*)) shown in (**A**) are additionally used to check for a possible quasi-resonant ac Zeeman shift by fitting a linear model and extrapolating to 0 mW. The fit (dashed blue line) implies no significant shift. The zero-field $F$=0 frequency and uncertainty are indicated by the red horizontal line and pink bands. (**D**) $F$=1 line-center frequencies from the fits shown in (**B**), after correction for systematic shifts (*22*). The purple line and bands indicate the weighted mean and uncertainty.

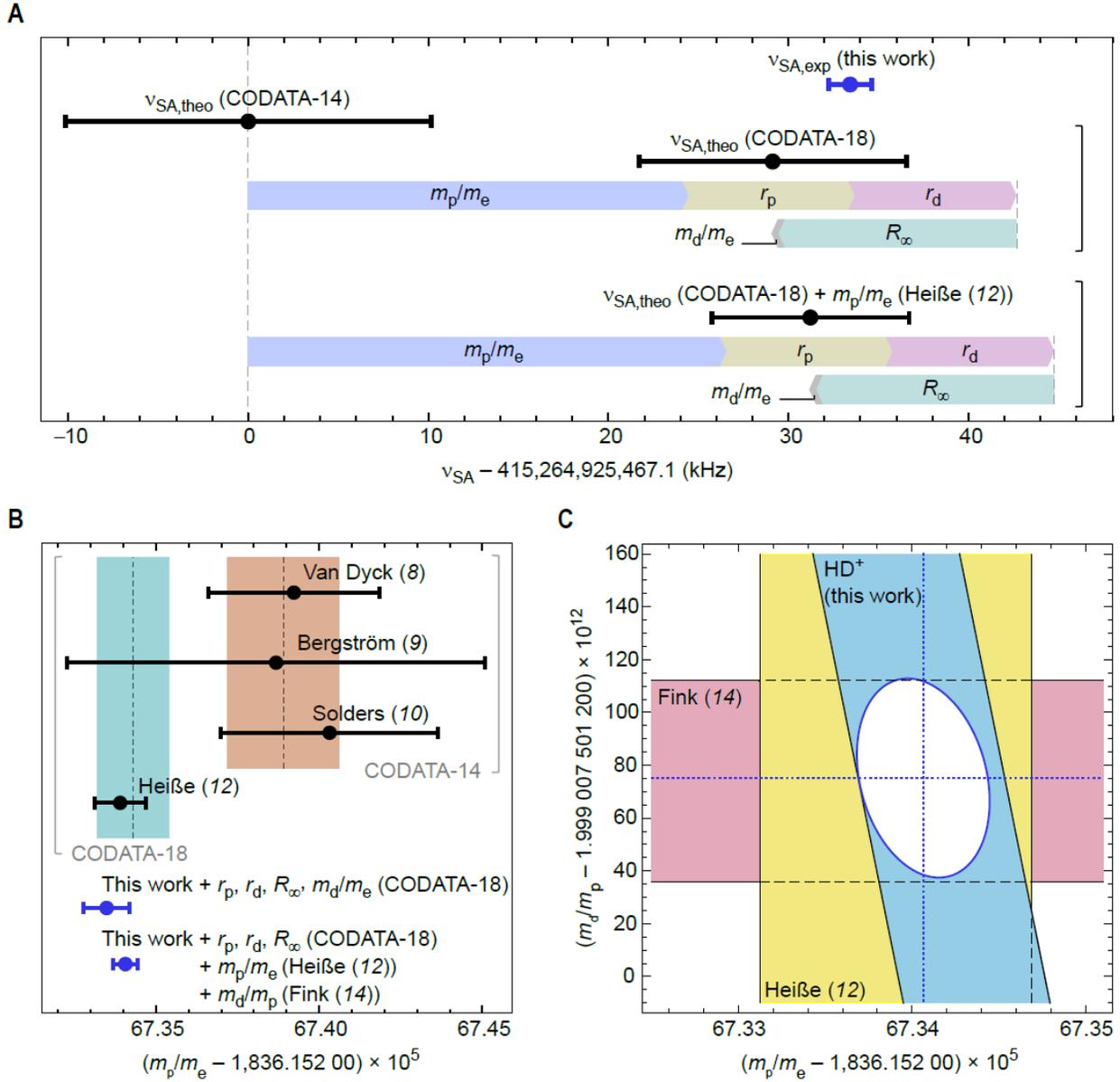

**Fig. 3. Implications for the values of physical constants.** (**A**) Comparison between $\nu_{SA,exp}$ and theoretical frequencies $\nu_{SA,theo}$ ($k$) obtained for the indicated combinations of fundamental constants, $k$. Arrows represent the cumulative frequency shift introduced by consecutively replacing the CODATA-2014 values of $m_p/m_e$ (blue), $r_p$ (yellow), $r_d$ (red), $R_\infty$ (green), and $m_d/m_e$ (gray), with their counterparts of the set $k$. (**B**) Values and uncertainties of $m_p/m_e$ from this work (blue data points) compared with measured $m_p$ values from other sources, which were converted

to values of $m_p/m_e$ through division by $m_e$ (CODATA-2018). The bottommost blue data point represents the value derived in (**C**). Dashed lines and shaded areas represent CODATA values and their $\pm 1\sigma$ ranges, with brackets indicating which of the measurements shown were included in the respective CODATA adjustments. (**C**) Simultaneous constraint on $m_p/m_e$ and $m_d/m_p$ from $HD^+$ and recent independent measurements of these quantities, leading to new values of $m_p/m_e$ and $m_d/m_p$, indicated by the blue dotted lines, and the corresponding $1\sigma$-constrained region indicated by the white ellipse.

**Table 1. Leading systematic shifts and uncertainties.** Shifts and their standard uncertainties (within parentheses) are given in kHz. Their justification can be found in (*22*), as well as the complete error budget (Table S2).

| Description | *F*=0 transition | *F*=1 transition |
|---|---|---|
| dc Zeeman effect | 0.02(1) | 0.10(1) |
| ac Stark effect 532 nm laser | 0.41(10) | 0.46(11) |
| ac Stark effect 1442 nm laser | −0.06(1) | −0.01(0) |
| ac Stark effect 1445 nm laser | 0.03(1) | −0.11(3) |
| Atomic frequency reference and ultrastable laser drift | −0.02(42) | −0.02(42) |
| **Total systematic shifts** | **0.38(43)** | **0.42(43)** |
| Uncertainty of fitted optical transition frequencies | 0.00(41) | 0.00(51) |
| **Total systematic shifts + fitted optical frequencies** | **0.38(59)** | **0.42(66)** |

**Table 2. Experimental and theoretical transition frequencies and hyperfine intervals.** Uncertainties are given within parentheses, and justified in detail in (*22*). The uncertainties of hyperfine intervals include the expansion factor of about two. During data acquisition and in Fig. 2, theoretical frequency values $\nu_{HF0,theo}$ and $\nu_{HF1,theo}$ based on CODATA-2014 constants were used as offset values; these are included for completeness and labeled with an asterisk. All other theoretical frequency values are obtained using CODATA-2018 physical constants.

| Symbol | Value (kHz) |
|---|---|
| $\nu_{HF0,theo}$* | 415,265,040,466.8 |
| $\nu_{HF1,theo}$* | 415,264,862,219.1 |
| $\nu_{HF0,exp}$ | 415,265,040,503.6(0.6) |
| $\nu_{HF1,exp}$ | 415,264,862,249.2(0.7) |
| $f_{0c,theo}$ | 114,999.7(1.9) |
| $f_{1c,theo}$ | −63,248.0(2.1) |
| $f_{10,theo}$ | 178,247.7(3.3) |
| $f_{10,exp}$ | 178,254.4(0.9) |
| $\nu_{SA,theo}$ | 415,264,925,496.2(7.4) |
| $\nu_{SA,exp}$ | 415,264,925,500.5(1.2) |

# Supplementary Materials for

## Proton-Electron Mass Ratio from Laser Spectroscopy of HD$^+$ at the Part-Per-Trillion Level

Sayan Patra, M. Germann, J.-Ph. Karr, M. Haidar, L. Hilico, V. I. Korobov, F. M. J. Cozijn, K. S. E. Eikema, W. Ubachs, J. C. J. Koelemeij*

*Correspondence to: j.c.j.koelemeij@vu.nl

**This PDF file includes:**

Materials and Methods
Figs. S1 to S3
Tables S1 to S3
References

**Materials and Methods**

**1. Experimental procedure**

1.1. <u>Acquisition of spectral data</u>

The ion trap apparatus and detection method have been described in detail previously (*23,29*). In short, about 40 to 100 HD$^+$ molecular ions are stored near the symmetry axis of a linear Paul trap surrounded by a shell of 1000 to 1500 Be$^+$ ions, laser-cooled by a single beam of circularly polarized 313 nm radiation. Sympathetic cooling leads to the formation of so-called Coulomb crystals with the HD$^+$ ions having a (secular) motional temperature of about 10 mK. A measure of the number of HD$^+$ ions is obtained by driving the radial center-of-mass motional mode of the HD$^+$ core using a resonant ac electric field ('secular excitation'), such that the Coulomb crystal heats up and melts, resulting in a rise of the Be$^+$ 313 nm fluorescence level approximately proportional to the number of HD$^+$ ions. This capability is used to detect the fractional loss of HD$^+$ molecules following resonance-enhanced multiphoton dissociation (REMPD), which involves the 1442 nm and 1445 nm photons, as well as one 532 nm photon from a continuous-wave Nd:YVO$_4$ laser which dissociates HD$^+$ only if the $v$=9 level is populated (Fig. 1A). The two-photon signal thus manifests itself in the form of increased loss of HD$^+$.

One experimental cycle involves the following steps. From a mixed-species Coulomb crystal already present in the trap, all impurity ions (including HD$^+$ remaining from a previous cycle) are removed using standard techniques (such as mass-selective over-driving of the secular motion of ions lighter than Be$^+$ using resonant ac electric fields, and mass-selective spill-over of heavier ions due to a temporary static quadrupole potential applied to the trap electrodes) so that only Be$^+$ ions remain. HD$^+$ ions are created during about 1.5 s of electron-impact ionization of neutral HD molecules, and subsequently captured and sympathetically cooled by the Be$^+$ ions. A measure of the initial number of $HD^+$ ions is obtained by secular excitation spanning a 10–12 s interval, which is done at relatively high intensity and large red detuning of the 313 nm laser (*29*), and using a quantization magnetic field of typically 1 G aligned with the wave vector of the 313 nm laser. Next, the quantization magnetic field is reduced (see Sec. 2.2), and the 313 nm cooling laser parameters are tuned so as to reduce the (secular) motional temperature of the $HD^+$ ions to approximately 10 mK. Then, with the cooling laser on, the $HD^+$ ions are exposed to the 532 nm, 1442 nm and 1445 nm laser beams for a period of 30 s. During this period, loss of $HD^+$ occurs due to collisions and chemical reactions with background-gas constituents (primarily $H_2$), as well as REMPD. The REMPD signal contains contributions from sequential (Doppler-broadened) transitions as well as direct (Doppler-free) two-photon excitation (*21*). Afterwards, a second secular excitation is performed to obtain a measure of the remaining number of $HD^+$ ions, and combined with the result of the first one to estimate the cumulative fractional loss of $HD^+$ ions. The corresponding frequency of the 1442 nm and 1445 nm lasers is determined as described in the next section.

Since data acquisition for a single spectrum takes place over time spans of several days, a number of experimental measures were taken to ensure that relevant experimental conditions remain constant. These include magnetic field conditions (see Sec. 2.2), stray electric fields in the ion trap, geometric overlap of the laser beams with the trapped ions, as well as the numbers and temperature of trapped $Be^+$ and $HD^+$ ions. The latter two are verified through images of the 313 nm $Be^+$ fluorescence taken with an electron-multiplying charge-coupled device (EMCCD) camera (*23*). The EMCCD camera images of the mixed-species Coulomb crystal are also used to detect and compensate stray electric fields, whose presence is inferred from the radial displacement of the $Be^+$ ions relative to the more tightly confined $HD^+$ ions when the radial confinement is modulated. As a result, the mixed-species Coulomb crystal was always located at the same position relative to the 313 nm, 532 nm, 1442 nm and 1445 nm laser beams, whose positions were held constant by carefully overlapping and aligning them through diaphragms placed on fixed locations near the entry and exit viewports of the vacuum chamber. Of these overlapped laser beams, the 313 nm laser provided an online visual check on the degree of overlap with the Be$^+$ and HD$^+$ ions. From measured residual beam-pointing errors (*23*) and laser power variations we infer that the laser beam intensities stayed constant at the location of the ions to within 25% of their nominal values.

1.2. <u>Frequency measurement and stabilization of the 1442 and 1445 nm spectroscopy lasers</u>

The 1442 nm and 1445 nm spectroscopy lasers are commercial external cavity diode lasers (ECDLs), which are individually phase-locked to an ultrastable fiber laser at 1542 nm (stabilized to a reference cavity made of ultra-low expansion (ULE) glass, resulting in a line width below 2 Hz) using an optical frequency comb laser as a transfer oscillator (*30*). This technique involves the realization of a so-called virtual beat note between each ECDL and the ultrastable laser. For each of the 1442 nm and 1445 nm lasers, the virtual beat-note signal is mixed with the output of an rf signal generator, and subsequently sent through a low-pass filter. This provides an error signal to a high-bandwidth feedback loop, which stabilizes the optical phase of each ECDL to the phase of the 1542 nm laser. At the same time, optical frequencies are measured using the frequency comb laser, which is referenced to a commercial cesium atomic clock with a relative frequency uncertainty below 1 ppt. The optical outputs of the 1442 nm and 1445 nm ECDLs are transported to the ion trap through separate passive optical fibers of 40 m length, which results into laser line-width broadening to 1−2 kHz at the location of the ions.

During the measurements, the detuning of the 1442 nm laser frequency for the $F$=0 transition is held fixed at $\delta_0$ = +16.5 MHz with respect to the ($v$,$L$; $F$,$S$,$J$) : (0,3;0,1,4) $\rightarrow$ (4,2;0,1,3) transition. For the $F$=1 transition, $\delta_1$ = $-16.8$ MHz with respect to the ($v$,$L$; $F$,$S$,$J$) : (0,3;1,2,5) $\rightarrow$ (4,2;1,2,4) transition (Fig. 1B). With these choices for the detunings, excessive population of the intermediate $v$=4 state is avoided, thereby suppressing the background signal due to Doppler-broadened excitation while still preserving a sufficient direct two-photon excitation rate (*21*). Before each measurement cycle, we set the frequency of the 1445 nm laser to its desired value by adjusting the frequency of the rf signal generator (in steps of 2 kHz) which determines the frequency of the virtual beat note with the 1542 nm laser.

Effectively, both the 1442 nm and 1445 nm lasers are phase-locked to the 1542 nm laser. Over the course of the experiment, however, the frequency of the 1542 nm laser drifts slowly and linearly (at a long-term drift rate of +0.02 Hz/s) with respect to the cesium-clock-stabilized optical frequency comb laser, due to a slow length change of the ULE reference cavity. This drift is measured over time intervals of 1000 s, and a frequency correction is calculated and applied to the rf signal generators of both the 1442 nm and the 1445 nm lasers before each cycle. Because of the 1000 s averaging period and the duration of the experimental cycle itself, this frequency correction lags the actual laser drift by (on average) 530 s. Since this drift affects both the 1442 nm and 1445 nm laser frequencies, it effectively translates to a small two-photon laser frequency shift of 23 Hz, which we treat (and correct for) as a systematic offset. We furthermore verified experimentally that the actual (measured) optical frequencies of the 1442 nm and 1445 nm lasers are equal to the target (set) frequency values to within ±0.1 kHz, i.e. well within the 0.42 kHz frequency uncertainty due to the 1 ppt relative frequency uncertainty of the cesium reference clock.

### 1.3. Population increase through rf spin-flip transitions

During part of the data acquisition for the $F=0$ transition, the population of the $(v,L; F,S,J) = (0,3;0,1,4)$ initial state was increased by transferring population from the $(v,L; F,S,J) = (0,3;1,2,5)$ and $(v,L; F,S,J) = (0,3;1,2,4)$ hyperfine states through spin-flips induced by resonant rf magnetic fields with frequencies $f_{RF1} = 1050.81$ MHz and $f_{RF2} = 1037.26$ MHz, respectively (Fig. S1). These fields were generated by combining the signal outputs of a first rf generator oscillating at 1044.055 MHz and a second rf generator at 6.774 MHz using a mixer. The resulting signal was subsequently amplified and fed into a loop antenna placed at a distance of 5 cm from the ion trap center. The loop antenna's symmetry axis was oriented at an angle of roughly 45° with the quantization axis (defined by the quantization dc magnetic field), in order to drive all $\pi$, $\sigma^+$ and $\sigma^-$ spin-flip transitions. The second rf generator was modulated by noise with a bandwidth of 200 kHz, wide enough to cover the Zeeman shifts of all magnetic subcomponents of the spin-flip transitions.

## 2. Line shape model and line shifts

### 2.1. Hyperfine structure

The hyperfine structure of the $v=0$ and $v=9$ states is obtained from the eigenvalues of an effective spin Hamiltonian, derived from the Breit-Pauli Hamiltonian. The effective spin Hamiltonian captures various spin-spin interactions, expressed as the sum of nine tensor products of the angular momentum operators $\mathbf{L}$, $\mathbf{s}_e$, $\mathbf{I}_p$, and $\mathbf{I}_d$ (*24*). The strengths of the nine interaction terms are specified by proportionality coefficients, $E_i^{vL}$ (with $i \in \{1\ldots9\}$), which are computed for a given rovibrational state with vibrational quantum number $v$ and rotational quantum number $L$ (*24*). In this work, $L=3$ for both the $v=0$ initial state and $v=9$ final state, and in what follows we drop the superscript $L$ from $E_i^{vL}$. The precision of the largest coefficients ($E_4^v$, $E_5^v$) was recently increased by including corrections of order $\alpha^2$ and $\alpha^3$ (with $\alpha$ the fine-structure constant) relative to the Breit-Pauli terms, as well as nuclear-structure effects of the proton and deuteron (*26*), in the framework of nonrelativistic quantum electrodynamics (NRQED). $\alpha^2$-order corrections were also calculated for the third-largest coefficient, $E_1^v$. Our approach, which is similar to that used by Douglas and Kroll (*31*) and Pachucki (*32*) to calculate $\alpha^2$-order corrections to the helium fine structure, was recently successfully tested by applying it to the hydrogen atom, and comparing its predictions with the well-known hyperfine structure of atomic hydrogen (*33*). Numerical values of the spin coefficients are provided in Table S1.

We express the effective spin Hamiltonian in matrix form by computing matrix elements in the basis of 'pure' states (using ket notation) $|F,S,J,M_J\rangle$, with $M_J$ the eigenvalue of the operator $J_z$ (the projection of $\mathbf{J}$ on the space-fixed quantization axis)[1]. In this basis, the effective spin Hamiltonian matrix is almost diagonal, such that $F$ and $S$ are approximately good quantum numbers, while $J$ and $M_J$ are good quantum numbers. Diagonalization of the effective spin Hamiltonian matrix then yields all relevant hyperfine energies and intervals, including those indicated in Figs. 1B,C.

## 2.2. Zeeman effect

For a given vibrational state (with $v$=0 or 9), matrix elements of the Zeeman effect are evaluated for the pure states (*21*), and added to the effective spin Hamiltonian matrix. Diagonalization then yields the hyperfine structure as well as the Zeeman shift, which for each eigenstate is fitted with a second-order polynomial as done in (*34*) in order to evaluate separately the contributions by static (linear plus quadratic Zeeman shift) and quasi-static ac magnetic fields (quadratic Zeeman shift only). The Zeeman effect is dominated by the electron spin, which contributes a relatively large linear Zeeman shift of order 1 MHz/G. However, by selecting transitions between upper and lower states which have identical quantum numbers ($F,S,J,M_J$), the Zeeman effect in the two-photon transition is strongly suppressed (*21*). For transitions with $\Delta M_J = 0$, the residual linear Zeeman shift is less than 90 Hz/G for the $F$=1 transition, and below 47 kHz/G for the $F$=0 transition.

---

[1] Following established notation for the HD$^+$ molecular ion, we denote the total angular momentum quantum number as $J$, rather than $F$ as in atoms and other molecules.

We employ static quantization magnetic fields (directed along the 313 nm cooling laser beam) of typically −114(7) mG for the $F$=1 transition, and 18(7) mG for the $F$=0 transition, which has a larger quadratic Zeeman effect. Transverse static magnetic fields are nulled using orthogonal pairs of shim coils by minimizing optical pumping out of the 313 nm cycling transition (equivalent to optimizing the brightness of the fluorescing Be$^+$ ions as seen on the EMCCD camera). With the transverse fields shimmed out, the magnitude of the field is determined in two steps. First, the magnetic field zero is found by adjusting the electric current through the quantization field coils such that the Be$^+$ fluorescence drops to a sharp minimum (indicating optical de-pumping caused by misalignment of the quantization axis due to small residual transverse fields). From there, the current is adjusted until the desired quantization field is achieved (employing known current-to-field conversion factors). It was verified that transverse fields remain shimmed out throughout this procedure, and the procedure was repeated a few times per day of measurement. This also provides statistics on the reproducibility of the magnetic field settings, which are included in the magnetic-field uncertainty.

In the ion trap apparatus also ac magnetic fields are present, oscillating at several frequencies ranging from 50 Hz to 1 kHz. These were measured with a NIST-traceable magnetic field probe, with the strongest frequency components having an amplitude of 32(10) mG with spatial field components perpendicular to the quantization magnetic field. Such fields can in principle drive transitions between different $M_J$ states, and thereby change the spectral line shape. To assess this, we solve the time-dependent Schrödinger equation, assuming a mixed (thermal) initial molecular state having equal $M_J$ populations and random phases as the result of the interaction of the molecule's rotational states with ambient blackbody radiation (BBR). In this case, all $M_J$ states remain equally populated, leading to no deformation of the line shape by optical pumping, and leaving only the small quadratic Zeeman shift.

## 2.3. Stark effect

Similar to the Zeeman effect, matrix elements of the Stark effect are evaluated for the pure states (*25,35*). This is done for each of the optical fields at 313 nm, 532 nm, 1442 nm, and 1445 nm. For the quasi-resonant lasers at 1442 nm and 1445 nm, we take ion-motion and finite-lifetime effects into account as prescribed by (*36*). Together with the Zeeman Hamiltonian matrix, the resulting Stark Hamiltonian matrix is added to the effective spin Hamiltonian matrix prior to diagonalization. The 300 K BBR field and the rf electric field of the trap do not contribute significantly to the Stark effect at the 0.01 kHz level (*35,37*).

Although the various $M_J$ components of the $F$=0,1 transitions undergo shifts due to the Zeeman and Stark effects, these are smaller than the laser line width, and individual $M_J$ components are not resolved (Fig. S2). For the $F$=0 transition, the weak $\Delta M_J = \pm 2$ Zeeman satellite lines are encompassed by the power-broadened line width (Figs. S2A,B).

## 2.4. Two-photon line shape model

For each state with $v$=0 or $v$=9, diagonalization of the combined effective spin, Zeeman and Stark Hamiltonian matrix produces the hyperfine energies including the Zeeman effect (dependent on $M_J$) and Stark effect (dependent on $|M_J|$) caused by the magnetic and laser electric fields present in the trap. We calculate Rabi frequencies for each of the hyperfine magnetic subcomponents of the ($v$=0,$L$=3; $F,S,J$, $M_J$) → ($v$=4,$L$=2; $F,S,J$−1,$M_J$') and ($v$=4,$L$=2; $F,S,J$−1, $M_J$') → ($v$=9,$L$=3; $F,S,J$, $M_J$'') transitions, with ($F,S,J$) = (0,1,4) or (1,2,5) (*23,37*). From these we obtain two-photon transition rates and power broadening factors, which are further combined with the dissociation rate to obtain the full power- and interaction-time-broadened line width (*21*). With the 1442 nm and 1445 nm lasers being linearly polarized and propagating in parallel with the quantization axis, the fields can be seen as equal superpositions of left-handed and right-handed circularly polarized light, and the lasers primarily drive σ⁺ and σ⁻ transitions, leading to the selection rule $\Delta M_J = M_J'' − M_J = 0, \pm2$. Small residual misalignments between the 1442 nm and 1445 nm lasers, and a possible small degree of birefringence introduced by the vacuum chamber's viewports, may lead to some depolarization (*23*). This could result into a small imbalance in the rates at which σ⁺ and σ⁻ transitions are driven, as well as a small contribution by π transitions to the spectrum, with corresponding selection rule $\Delta M_J = 0, \pm1$. All these effects are parametrized and taken into account through the state of polarization of the 1442 nm and 1445 nm lasers, such that parameter uncertainty margins (residual misalignment, birefringence) can be propagated through the line-shape model to estimate the resulting line shift.

The two-photon transition line shape is approximated by a sum of Lorentzian line shapes, one for each magnetic subcomponent of the transition, with each line characterized by its (hyperfine-, Zeeman-, and Stark-shifted) line center, its full-width-half-maximum line width (due to the 1−2 kHz laser line width, 1−10 kHz of $M_J$-dependent power broadening, and 1.4 kHz of interaction-time broadening), and transition strength. As a result of Zeeman and Stark spreading of the individual magnetic subcomponents, the line shape may become slightly asymmetric for the fields present in our setup. Example line shapes indicating typical shifts and line-shape deformations are shown in Figs. S2B,D.

In the line-shape model, we assume that each magnetic substate of the $v$=9 excited states is dissociated by the 532 nm laser before it decays by spontaneous emission. This is justified given the state of polarization of the 532 nm laser (leading to $\sigma^+$ and $\sigma^-$ transitions driven at equal rates), and given that the average dissociation-limited lifetime of magnetic substates is 0.2 ms, much shorter than the spontaneous lifetime of about 12 ms. In addition, spontaneous emission (if it occurs) will most often proceed in a stepwise manner, taking the $HD^+$ molecules through $v$=8 and $v$=7 states which are also efficiently dissociated by the 532 nm laser ([29]).

## 2.5. Other systematic effects and size of systematic shifts

Apart from the systematic effects described above, a number of other, smaller effects are evaluated. These include collisional shifts, the electric quadrupole shift ([38]), the second-order Doppler shift ([23]), and the shift due the background slope caused by the Doppler-broadened two-photon signal. The collisional frequency shift is obtained by multiplying the Langevin spiraling collision rate, given the $1 \times 10^{-8}$ Pa molecular-hydrogen background pressure in our apparatus, with the worst-case phase shift of $\pi/2$ which could arise from such a collision (about one collision per 30 s). This leads to a maximum frequency shift of 8 mHz. The maximum shift (2 Hz) due to the Doppler-broadened background slope is estimated from an Einstein rate-equation model ([23]) used to predict the combined Doppler-broadened and Doppler-free two-photon spectrum ([39]).

Another line-shifting mechanism to consider is the so-called cross-damping or quantum interference ([40]); see also ([41]) and references therein. This effect comes into play when the excited-state population is determined through the detection of photons that are spontaneously emitted following resonant excitation by the spectroscopy laser. In some cases this laser may off-resonantly excite other states, opening up additional spontaneous emission channels which interfere with the primary decay channel, thus leading to an apparent shift of the excitation spectrum. We have considered the possible presence of such shifts in our signals. In our case, however, we do not detect fluorescence photons, but instead determine excited-state populations through (incoherent) photodissociation by the 532 nm laser and measuring the subsequent loss of $HD^+$ ions. For this reason, effects of cross-damping are ruled out.

It is instructive at this point to compare the sizes of the various sources of uncertainty in our work. The theoretical uncertainty margin (3.1 kHz) and the uncertainty contributed by the physical constants (5-10 kHz) are far greater than the statistical precision (0.6-0.7 kHz) of our spectroscopic method. Furthermore, Table S2 shows that all systematic shifts (and their uncertainties) are small, and at most comparable in size to the statistical precision. Experimental determination of systematic shifts is therefore challenging, and instead we follow a different approach to account for systematic shifts. With the line-shape model described above, we can simulate spectra for experimentally determined external field parameters (*e.g.* magnetic fields, laser intensities). These simulated line shapes thus include the corresponding line shift and line deformation, and may furthermore be extended to include measurement noise and Poisson noise caused by the small numbers of $HD^+$ ions being dissociated. Next, we fit a simple Lorentzian line shape function to the simulated line shape, analogous to how Lorentzian line shapes are fitted to the experimental spectra. Taking the difference between the fitted line center and the field-free transition frequency then yields the total shift to be applied to the measured frequency value. We furthermore note that this procedure takes into account pure line shifts as well as shifts due to line-shape deformation.

To obtain insight in the shift caused by each effect alone, we also evaluate each systematic shift separately. We do this by first fitting a Lorentzian line shape function to the simulated spectrum for the nominal values of all experimental parameters, followed by another Lorentzian fit to a second simulated spectrum, this time with the parameter of interest set to zero. The shift is then found from the difference between the fitted Lorentzian line centers. For each of the ac Stark shifts caused by the 532 nm, 1442 nm and 1445 nm lasers, we use a slightly different approach, as setting these laser intensities to zero leads to a vanishing signal. In this case, we determine the frequency shifts for 75%, 50%, and 25% of the nominal intensity, and extrapolate to zero intensity to find the total shift. The uncertainty caused by each systematic effect is found in a similar way, by comparing fitted Lorentzian line centers found for nominal values of external field parameters, with line centers found in the case where the parameter of interest was increased by its (measurement) uncertainty. The thus found uncertainties are included in the evaluation of the overall experimental uncertainty. All shifts and uncertainties are listed in Table S2.

Such an assessment of systematic shifts is only meaningful if both the molecular response to an external field and the field itself are sufficiently well known (as is the case for the Zeeman and Stark effect, magnetic fields, laser intensities, and polarization vectors). For the quasi-resonant rf magnetic fields produced by the loop antenna, however, this is not the case. Although we know *a priori* that the shift due these fields should vary proportionally to the rf power, we lack precise information about the efficiency of the antenna and the attenuation of the rf field by the ion trap structure, which prevents us from estimating the value of the frequency shift with sufficient accuracy. In this case, and as further detailed in Sec. 2.6 below, we measure the $F$=0 transition frequency for various values of the rf power, $P_{rf}$, sent into the loop antenna. We correct each transition frequency for the well-understood systematic shifts, and subsequently fit a linear function proportional to $P_{\text{rf}}$ to the corrected frequency data, and extrapolate to zero power to find the zero-field frequency (Fig. 2C).

2.6. <u>Experimental checks on systematic effects.</u>

In addition to the above-mentioned theory-based estimation of systematic effects, we conducted a number of tests to experimentally verify their small size and relative insignificance. Of the various ac Stark shifts, for example, by far the largest contribution stems from the 532 nm laser. Still, the maximum change in ac Stark shift achievable in our setup (~0.3 kHz) is smaller than the statistical precision of the fitted line centers (0.6 – 0.7 kHz). It is nevertheless useful to verify experimentally that the shift is as small as estimated theoretically, which in this case implies that no shift should be visible if the 532 nm laser intensity is varied. We have therefore measured both the $F$=0 and $F$=1 transition frequencies for 532 nm laser intensities of 2.5 MW/m$^2$ and 0.57 MW/m$^2$. The estimated change in ac Stark shift is −0.33 kHz for the $F$=0 transition, and −0.36 kHz for the $F$=1 transition. As expected, we observe no significant shift to within the ~1 kHz precision of the line-shift measurement: experimentally measured shifts are −0.1(1.0) kHz for $F$=0 and 0.2(1.1) kHz for $F$=1. All measured transition frequencies are individually corrected for the theoretically estimated ac Stark shift before calculating the final frequency values.

We furthermore varied the power of the rf magnetic field which was used to increase the population available for the $F$=0 transition (Secs. 1.3 and 2.5 above), using rf input power values of 0 mW, 5 mW (the nominal value), and 100 mW. A first $F$=0 spectrum was acquired at 0 mW (i.e. no rf applied), after which we recorded a second spectrum at 5 mW, which is the smallest rf power for which the signal-to-noise ratio is significantly increased. Both measurements produced the same transition frequency (within the statistical precision of the fitted line center). However, for such near-resonant rf fields we may expect a small near-resonant ac Zeeman effect, with a shift of the form $\Omega_{rf}^2/\Delta$. Here $\Omega_{rf}$ stands for the Rabi frequency (which depends on poorly known quantities such as the antenna efficiency and rf-field attenuation by the ion trap), and $\Delta$ takes into account the detuning from the spin-flip resonance frequency. Due to the frequency noise applied to the rf signal, $\Delta$ has a pseudo-random dependence on time, and the shift would have to be computed as the time average, $\langle\Omega_{rf}^2/\Delta\rangle$. To check if such a shift could be observed we recorded a third $F$=0 spectrum at 100 mW of rf power (the maximum power available in our experimental setup). Based on coarse estimates of $\langle\Omega_{rf}^2/\Delta\rangle$ for worst-case and best-case scenarios, we expect a shift with a magnitude between 0.005 kHz and 0.35 kHz at 5 mW of rf power, and between 0.1 kHz and 7.1 kHz at 100 mW of rf power. Also the 100 mW measurement revealed no significant frequency shift. To include all three measurements in the analysis, we fit the slope and offset of our power-dependent model to the three data points of Fig. 2C. We thus obtain an experimental (and essentially zero) value of the shift of 0.3(1.3) kHz at 100 mW (Fig. 2C), which is consistent with our estimated range (yet more precise). The fitted zero-power offset frequency is used as input value in the $F$=0 frequency analysis.

## 3. Hyperfine structure and experimental spin-averaged frequency

To find $\nu_{SA,exp}$, we may take the hyperfine intervals, $f_{0c,theo}$ and $f_{1c,theo}$ from theory (*24,26*), and then compute $\nu_{SA,exp}$ as the mean of the two values $\nu_{HF0,exp} - f_{0c,theo}$ and $\nu_{HF1,exp} - f_{1c,theo}$. Consequently, the uncertainty of $\nu_{SA,exp}$ will depend not only on the experimental uncertainty, but also on the uncertainty of the theoretical hyperfine intervals (or an upper limit thereto). This uncertainty is determined by the precision of the spin coefficients $E_i^\nu$ which, however, is difficult to specify in terms of standard uncertainty. *A priori* estimates of the precision are typically based on order-of-magnitude estimates of the neglected higher-order terms, complemented by the observed level of agreement with experimental data (if available). In this way, the uncertainty of the theoretical spin coefficients is estimated to be about 1 part-per-million (ppm) for $E_4$ and $E_5$, about 3 ppm for $E_1$, and about 100 ppm for the other coefficients which, because of their small size (Table S1), were evaluated to lowest order only (Sec. 2.1). It is worth noting that the spin coefficient $E_4$ (associated with the electron-proton spin-spin interaction) has a similar counterpart in theoretical models describing the hyperfine structure of $H_2^+$, which was successfully compared with available experimental data from $H_2^+$ hyperfine spectroscopy at the 1-ppm level (*26*).

By expanding the relevant theoretical hyperfine intervals in terms of small (linear) variations of the coefficients $E_i^\nu$, we can propagate their uncertainties, and estimate the uncertainties of the hyperfine intervals themselves. Relevant hyperfine intervals are $f_{0c}, f_{1c}, f_{10}$ ($= f_{0c} - f_{1c}$, see Fig. 1C), and the sum $\Sigma_{10} \equiv f_{0c} + f_{1c}$. The latter plays a role in the determination of $\nu_{SA,exp}$ and, in particular, its uncertainty, which can be seen as follows. Above, we defined $\nu_{SA,exp}$ $= (\nu_{HF0,exp} + \nu_{HF1,exp})/2 - (f_{0c,theo} + f_{1c,theo})/2$. However, the expressions $f_{0c,theo}$ and $f_{1c,theo}$ are strongly correlated as both depend on the coefficients $E_i^\nu$. So, rather than evaluating their individual uncertainties and adding them in quadrature, we write $\nu_{SA,exp} = (\nu_{HF0,exp} + \nu_{HF1,exp})/2 - \Sigma_{10}/2$, and evaluate the uncertainty of $\Sigma_{10}$ at once (by expanding it in terms of $E_i^\nu$ and propagating their uncertainties) to take these correlations into account.

To find an upper limit to the uncertainty margin of the theoretical hyperfine frequencies, we take advantage of the fact that the difference $\nu_{HF0,exp} - \nu_{HF1,exp}$ provides an experimental value of the hyperfine interval, $f_{10,exp}$, which can be compared with the value predicted by the effective spin Hamiltonian, $f_{10,theo}$. We obtain an upper limit to the uncertainty of $f_{10,theo}$ in terms of standard uncertainty by requiring that $f_{10,theo}$ and $f_{10,exp}$ are consistent, i.e. we take as a null hypothesis that the values of $f_{10,theo}$ and $f_{10,exp}$ are in agreement within their combined uncertainty, at a significance level of 0.05. Mathematically, this translates to the requirement that $|f_{10,exp} -$ $f_{10,theo}| < 1.96 \, \sigma_c$, with $\sigma_c = (\sigma_{10,exp}{}^2 + \sigma_{10,theo}{}^2)^{1/2}$ being the combined uncertainty of the two values. $\sigma_{10,exp}{}^2$ is obtained from the uncertainties of $\nu_{HF0,exp}$ and $\nu_{HF1,exp}$, which are computed individually for each transition by adding the uncertainty of the frequency measurement and all the systematic uncertainties in quadrature (Table S2), leading to $\sigma_{10,exp} = 0.9$ kHz. The comparison between the experimental and theoretical values furthermore yields that $f_{10,exp} - f_{10,theo}$ $= 6.7$ kHz. It then follows that the probability value ($p$-value) exceeds the significance level of 0.05 (and the null hypothesis is accepted) if $\sigma_{10,theo} = 3.3$ kHz. This value is about a factor of two larger than the value of $\sigma_{10,theo}$ obtained from propagating the coarse uncertainties of the spin coefficients $E_i{}^\nu$ mentioned above. The same factor of two also applies to the uncertainty of $\Sigma_{10}$, such that we arrive at a value $\nu_{SA,exp} = 415,264,925,500.5(0.4)_{exp}(1.1)_{theo}$ kHz, with a combined uncertainty of 1.2 kHz. Table S3 provides an overview of the measured transition frequencies, relevant hyperfine intervals, and their final uncertainties.

From Table S3 it is clear that the theoretical quantities $\Sigma_{10}$ and $f_{10,theo}$ have quite different uncertainties, 2.2 kHz and 3.3 kHz, respectively, despite the fact they are both composed of $f_{0c,theo}$ and $f_{1c,theo}$ in equal amounts. This difference is explained by the aforementioned correlation between $f_{0c,theo}$ and $f_{1c,theo}$. In addition, only half of the uncertainty of $\Sigma_{10}$ (1.1 kHz) contributes to the uncertainty of $\nu_{SA,exp}$. The found value of $\nu_{SA,exp}$ is therefore considerably less dependent on variations of the theoretical hyperfine structure than the interval $f_{10,theo}$.

We note that also the experimental quantities $\nu_{HF0,exp} + \nu_{HF1,exp}$ (used in the determination of $\nu_{SA,exp}$) and $\nu_{HF0,exp} - \nu_{HF1,exp}$ (used in the determination of $f_{10,exp}$) involve systematic effects which may be correlated. This is the case for the Zeeman shift, whose uncertainty could involve an unknown bias in the magnetic field measurement common to both the $F{=}0$ and $F{=}1$ measurements. Although this correlation plays a role at the sub-0.01 kHz level only, we here take the Zeeman uncertainties as maximally correlated.

## 4. Connection with values of physical constants

### 4.1. CODATA-2014 and CODATA-2018 adjustments

The $3\sigma$ discrepancy between the CODATA-2014 atomic proton-mass value and that of (*11*) was addressed in the 2017 special CODATA update (*2*). The new value of (*11*) was included after expanding the uncertainties of the two dominant input data by a common factor of 1.7, so as to reduce the absolute value of the residuals below two. The same procedure also appears to have been followed for the CODATA-2018 set of adjusted values. While the detailed description of the adjustment is still being awaited, the CODATA-2018 values have been published online very recently (*42*). The new CODATA value of $m_p/m_e$ is located in between the CODATA-2014 value on the one hand, and our value and that of (*11*) on the other, but closer to the latter two (Fig. 3B). Note that very recently, the authors of (*11*) published an updated value of $m_p$, following a re-analysis of the systematic effects (*12*). This new value, which is shifted towards larger mass by $0.45\sigma$, is used in the present analysis.

It should furthermore be noted that the CODATA-2018 values of $R_\infty$, $r_p$, and the deuteron charge radius, $r_d$, now virtually coincide with the values from muonic hydrogen, albeit with larger uncertainty margins. Using the CODATA-2018 set of physical constants, a theoretical frequency of $\nu_{SA,theo}$ (CODATA-2018) = 415,264,925,496.2(7.4) kHz is obtained (see Fig. 3A, Table 2, and Table S3). While this value is consistent with our experimental value, we point out that it is still influenced by the inconsistent input data concerning the proton atomic mass.

## 4.2. Impact of physical constants on the theoretical frequency and comparison with experiment

We use the results of *ab initio* molecular energy level calculations to parametrize the transition frequency in terms of $m_p/m_e$ as well as $m_d/m_e$, $\alpha$, $R_\infty$, $r_p$, and $r_d$ (denoting the deuteron-electron mass ratio, fine-structure constant, Rydberg constant, proton charge radius, and deuteron charge radius, respectively). For the small variations of physical constants considered here, we can write down a linearized expression for $\nu_{SA,theo}$ in terms of the dimensionless sensitivity coefficients given in (*17*). Contrary to (*17*), however, we do not linearize the dependence on $r_p$ and $r_d$, but instead take into account the full nuclear structure correction term, which is quadratic in $r_p$ and $r_d$ (*27*). This allows us to accommodate the large (4%) discrepancy between CODATA-2014 and the muonic hydrogen values with sufficient precision. Furthermore, in part of our analysis we use different sensitivity coefficients for the mass ratios than those given in (*17*). As previously explained by us (*43*), we can choose between a parametrization of $\nu_{SA,theo}$ in terms of the mass ratios $m_p/m_e$ and $m_d/m_e$, or the ratios $m_p/m_e$ and $m_d/m_p$. As a rule of thumb, we choose to work with the deuteron-to-particle ratio which is known to highest precision. In (*17*) a parametrization in terms of $m_d/m_p$ was chosen (leading to the sensitivity coefficients derived there), which is appropriate if we want to compare our result with the very recent precise value of $m_d/m_p$ obtained by Fink and Myers (*14*); see Fig. 3C. On the other hand, when considering the CODATA-2018 set of constants, $m_d/m_e$ is more precisely known than $m_d/m_p$, which calls for a parametrization in $m_d/m_e$ if conclusions about CODATA-2018 constants are to be drawn. This parametrization was also used throughout the main text and figures, with the only exceptions being Fig. 3C, and the experimental value $m_p/m_e = 1{,}836.152\ 673\ 406(38)$ computed using the value of $m_d/m_p$ from (*14*), shown also as the bottommost data point of Fig. 3B.

For the parametrization in terms of $m_d/m_e$, we derive sensitivity coefficients for variations of $m_p/m_e$ and $m_d/m_e$ equaling $-0.2348$ and $-0.1175$, respectively (*43*); see also Sec. 4.3 below. Also used here is the sensitivity coefficient for variations in the fine-structure constant, $1.5 \times 10^{-5}$, although its contribution to the uncertainty is negligible (at the level of 1 Hz). Throughout the text and in the figures, the dependence on the fine-structure constant is therefore ignored.

We thus obtain fully parametrized expressions for $\nu_{SA,theo}$ through which we propagate the values and uncertainties of physical constants, as well as their covariances (obtained from the CODATA adjustment, if applicable). Results are shown in Fig. 3A. Uncertainties and correlations are treated in the same way as in the CODATA adjustments (*44*), and correlations turn out to play a non-negligible role. For example, when using the CODATA-2014 values, a simple quadratic sum of the individual contributions (including the theoretical precision of 3.1 kHz) yields an uncertainty of 10.6 kHz, compared to the 10.2 kHz uncertainty obtained if correlations are taken into account. One set of particularly strong correlations (between $R_\infty$, $r_p$, and $r_d$) stems from the very precise measurements of the $1S$–$2S$ transition (*45*) and the isotope shift (*46*) in atomic hydrogen and deuterium, leading to correlation coefficients close to unity. This correlation is also visible in our value of $\nu_{SA,theo}$: when shifting from CODATA-2014 to CODATA-2018 values, the shift due to nuclear-structure effects is 18.2 kHz, which is partly compensated by the shift due to $R_\infty$ ($-13.2$ kHz) to produce a net shift of 5.1 kHz (Fig. 3A).

### 4.3. Determination of $m_p/m_e$, $m_d/m_p$, and the atomic mass difference $m_p + m_d - m_h$

Strictly speaking, the vibrational frequencies of HD$^+$ depend on the reduced mass of the two nuclei relative to the electron mass, and a single measurement will yield the sum $m_e/m_p + m_e/m_d = m_e/m_p (1 + m_p/m_d)$. Therefore, to determine $m_p/m_e$ we need to fix the value of $m_d/m_e$ (or $m_d/m_p$, as discussed in Sec. 4.2) for which we proceed as follows. We start by noting that the uncertainty of $v_{SA,theo}$ is predominantly limited by the uncertainty of $m_p/m_e$ (Fig. S3), implying that this is the correct parameter to constrain. For the determination of $m_p/m_e$, we equate the parametrized expression of $v_{SA,theo}$ to $v_{SA,exp}$ to obtain a so-called observational equation (*44*). We can solve this equation for any of the parameters involved, and in this case we solve for $m_p/m_e$. From the solution, the corresponding value and uncertainty of $m_p/m_e$ are found by inserting the values of all other parameters (including the theoretical and experimental frequency values), and propagating their uncertainties and covariances (*44*). This leads to the results shown in Fig. 3B. When comparing our value of $m_p/m_e$ to any of the $m_p$ measurements from (*8-12*), we first convert these measurements to values of $m_p/m_e$ by dividing them by the CODATA-2018 value of $m_e$.

We may also choose to leave more than one parameter in the observational equation variable, as was done in Fig. 3C, which displays a contour plot in the ($m_p/m_e$, $m_d/m_p$) plane, showing the [−1σ,1σ] constrained region derived from our measurement. This can be plotted together with the bounds obtained from individual $m_p/m_e$ (*12*) and $m_d/m_p$ (*14*) determinations. Assuming bivariate normal probability density distributions for each region, the total (product) probability density distribution for the three constrained regions together can be constructed, from which combined values of $m_p/m_e$ and $m_d/m_p$, as well as the 1σ-uncertainty ellipsoid, may be obtained (Fig. 3C).

From Fig. 3C we find values $m_p/m_e$ = 1,836.152 673 406(38) and $m_d/m_p$ = 1.999 007 501 275(38), with the latter being determined to a large extent by, and nearly equal to, the value from (*14*). The precision of the value of $m_p/m_e$, 21 ppt, is determined primarily by our present result and that from (*14*): if the value of $m_p/m_e$ from (*12*) is not taken into account, the precision deteriorates only slightly, to 24 ppt.

**Supplementary figures**

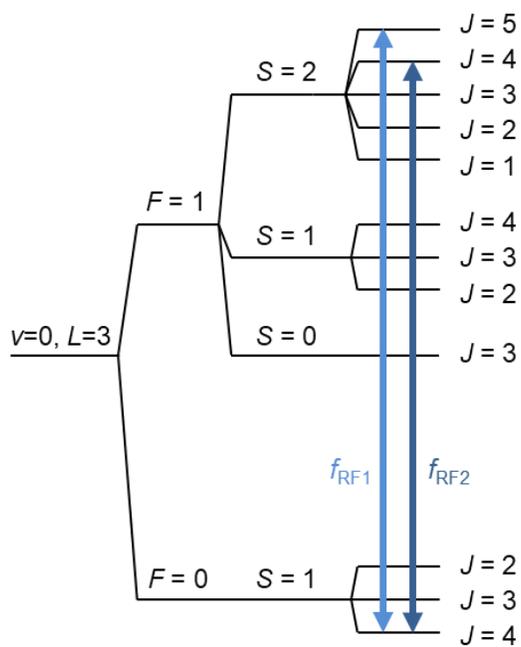

**Fig. S1. Partial level diagram (not to scale) and rf spin-flip transitions.** Two rf magnetic fields with frequencies $f_{RF1}$ = 1050.81 MHz and $f_{RF2}$ = 1037.26 MHz are generated, and modulated by noise with 200 kHz bandwidth so that they address all magnetic subcomponents of the indicated transitions.

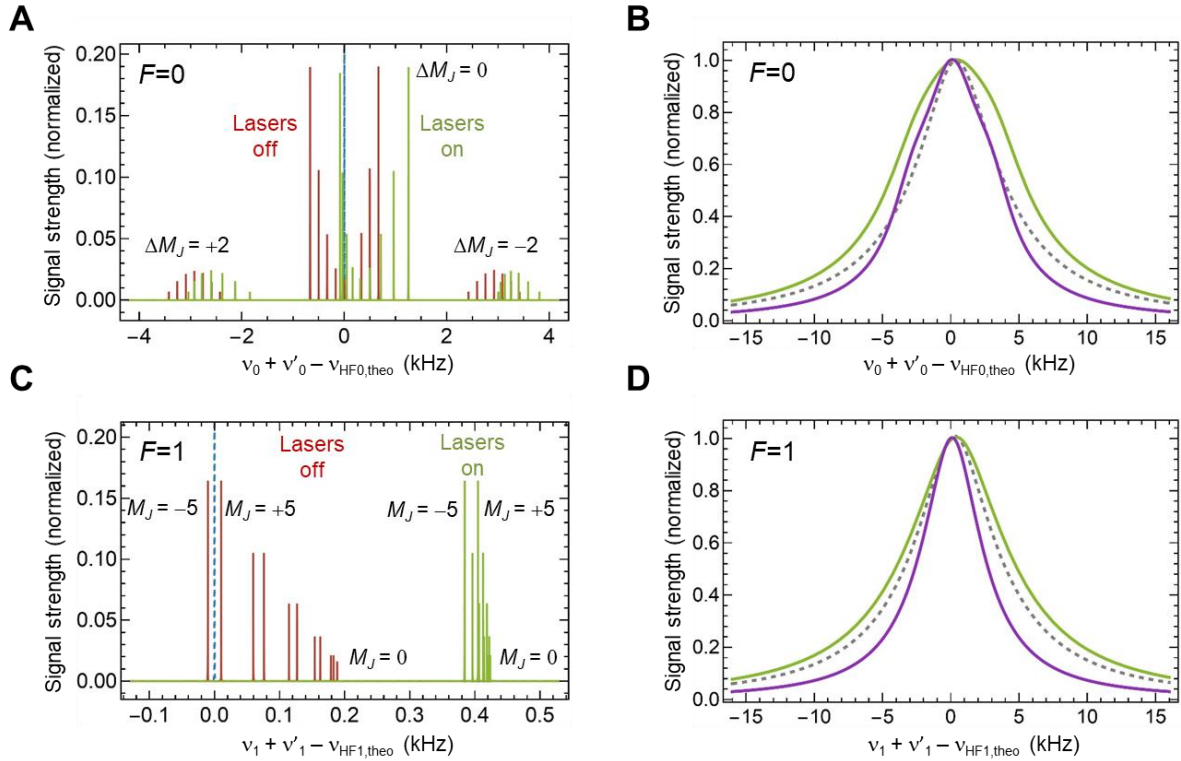

**Fig. S2. Line shifts and line deformation.** (**A**) Stick spectrum showing the relative signal strengths of magnetic subcomponents of the $F=0$ transition with $\Delta M_J = 0, \pm 2$. The blue dashed line at 0 kHz represents the collapsed stick spectrum in the absence of any magnetic or electric fields (with the line extending vertically to a peak value of 1). The red lines represent the Zeeman-shifted stick spectrum in the presence of a static magnetic field of 18 mG and a quasi-static ac field with amplitude 32 mG. For the green lines, also the ac Stark shift caused by all laser fields combined (at their maximum intensities) is included, while assuming the same magnetic field as for the red lines. (**B**). Fully broadened spectra for the $F=0$ transitions. Gray (dashed), zero magnetic field, maximum laser intensities; green, magnetic fields as described under (**A**) and maximum laser intensities; purple, same magnetic fields and laser intensities except for the 532 nm laser, which has an intensity reduced from 2.5 MW/m$^2$ to 0.57 MW/m$^2$. Zeeman broadening is visible when comparing the green to the gray dashed curve, while different levels of interaction-time broadening are visible in the green and purple curves.

(**C**) Stick spectrum showing the relative signal strengths of magnetic subcomponents of the $F$=1 transition with $\Delta M_J = 0$, and identifying the transitions between states with $M_J = \pm 5$ and $M_J = 0$. The $\Delta M_J = \pm 2$ satellite lines (not in view) are located at $\pm 60$ kHz. The blue dashed line again represents the collapsed stick spectrum in absence of any magnetic or electric fields (line extends vertically to a peak value of 1). The red lines represent the stick spectrum in the presence of a static magnetic field of $-114$ mG and a quasi-static ac field with amplitude 32 mG. The same magnetic field conditions apply to the green lines, which also include the Stark shift due to all laser fields (maximum intensities). (**D**) Fully broadened spectra for the $F$=1 transitions. Gray (dashed), zero magnetic field, maximum laser intensities; green, magnetic fields as described under (**C**) and maximum laser intensities; purple, same magnetic fields and laser intensities except for the 532 nm laser, which has an intensity reduced from 2.5 MW/m$^2$ to 0.57 MW/m$^2$.

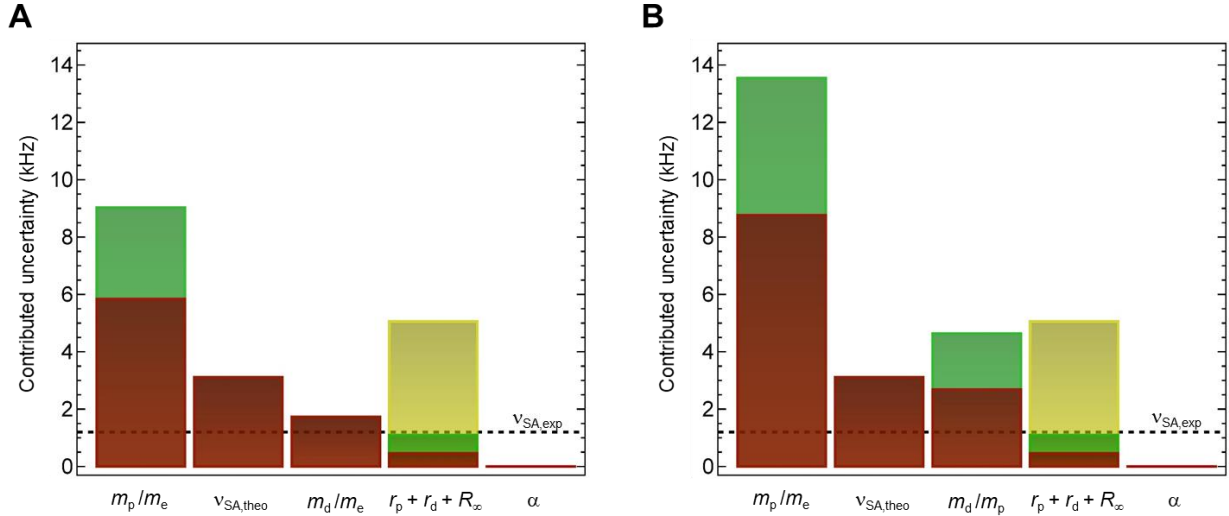

**Fig. S3. Uncertainty budget of $\nu_{SA,theo}(k)$.** Individual uncertainty contributions of physical constants and theory to the total uncertainty of $\nu_{SA,theo}$, here evaluated for $k$=CODATA-2014 (green bars), and for $k$=CODATA-2018 (red bars). (**A**) Uncertainties calculated using the parametrization in terms of $m_d/m_e$ (*43*). The yellow bar indicates the impact of the Proton Radius Puzzle for comparison. (**B**) Same as (**A**), but now using the parametrization in terms of $m_d/m_p$. In all cases, the largest contribution to the uncertainty stems from $m_p/m_e$, making this the correct parameter to constrain. As the individual contributions by $r_p$, $r_d$ and $R_\infty$ are strongly correlated through the $1S$–$2S$ transition frequency (*45*) and the isotope shift (*46*) measured in atomic hydrogen and deuterium, the graph shows the net uncertainty contributed by these three constants together. The dashed horizontal line represents the uncertainty of $\nu_{SA,exp}$, indicating that the precision of the comparison between theory and experiment is limited primarily by the precision of the value of $m_p/m_e$, as well as by theory.

**Supplementary tables**

**Table S1.**

**Spin coefficients of the effective spin Hamiltonian.** Numerical values of the spin coefficients $E_1$ through $E_9$ for $v$=0, $L$=3 and $v$=9, $L$=3. Units are kHz. For their definition and use in the effective spin Hamiltonian, see (*24,26*).

|        | $v$=0, $L$=3 | $v$=9, $L$=3 |
|--------|-------------:|-------------:|
| $E_1$  | 31,628.23    | 18,270.94    |
| $E_2$  | −30.83       | −21.30       |
| $E_3$  | −4.73        | −3.23        |
| $E_4$  | 920,481.65   | 775,706.33   |
| $E_5$  | 141,533.32   | 119,431.96   |
| $E_6$  | 948.52       | 538.99       |
| $E_7$  | 145.59       | 82.72        |
| $E_8$  | −0.33        | −0.22        |
| $E_9$  | 0.61         | 0.50         |

**Table S2.**

**Overview of systematic shifts.** Standard uncertainties are stated within parentheses. All uncertainties are evaluated around the nominal operation conditions. Shifts labeled with an asterisk are not taken automatically into account by the Lorentzian fit method described in Sec.

2.5, and are therefore subtracted separately. Shifts and uncertainties smaller than 0.005 kHz are rounded down to zero.

| Description | $F=0$ transition (kHz) | $F=1$ transition (kHz) |
|---|---|---|
| dc Zeeman effect | 0.02(1) | 0.10(1) |
| ac Zeeman effect | 0.00(0) | 0.00(0) |
| ac Stark effect 313 nm laser | 0.00(0) | 0.00(0) |
| ac Stark effect 532 nm laser | 0.41(10) | 0.46(11) |
| ac Stark effect 1442 nm laser | −0.06(1) | −0.01(0) |
| ac Stark effect 1445 nm laser | 0.03(1) | −0.11(3) |
| dc Stark effect trap and BBR electric field* | 0.00(0) | 0.00(0) |
| State of polarization of 1442 nm and 1445 nm lasers | 0.00(0) | 0.00(0) |
| Atomic frequency reference and ultrastable laser drift* | −0.02(42) | −0.02(42) |
| Electric quadrupole shift* | 0.00(0) | 0.00(0) |
| Slope due to Doppler-broadened background* | 0.00(0) | 0.00(0) |
| $2^{nd}$-order Doppler shift* | 0.00(1) | 0.00(1) |
| Collisional shift* | 0.00(0) | 0.00(0) |
| **Total systematic shifts** | **0.38(43)** | **0.42(43)** |
| Fitted optical transition frequencies | 0.00(41) | 0.00(51) |
| **Total systematic shifts + fitted optical frequencies** | **0.38(59)** | **0.42(66)** |

**Table S3.**

**Overview of experimental and theoretical transition frequencies and hyperfine intervals.**
Uncertainties are stated within parentheses. Theoretical values are obtained using CODATA 2018 physical constants. The uncertainties of the theoretical hyperfine intervals include the expansion factor described in Sec. 3.

| Description | Symbol | Value (kHz) |
|---|---|---|
| Measured $F=0$ transition frequency | $\nu_{HF0,exp}$ | 415,265,040,503.6(0.6) |
| Measured $F=1$ transition frequency | $\nu_{HF1,exp}$ | 415,264,862,249.2(0.7) |
| Experimental spin-averaged frequency | $\nu_{SA,exp}$ | 415,264,925,500.5(1.2) |
| Theoretical spin-averaged frequency | $\nu_{SA,theo}$ | 415,264,925,496.2(7.4) |
| Experimental $F=0$ hyperfine shift | $f_{0c,exp}$ | 115,003.0(0.6) |
| Theoretical $F=0$ hyperfine shift | $f_{0c,theo}$ | 114,999.7(1.9) |
| Experimental $F=1$ hyperfine shift | $f_{1c,exp}$ | −63,251.4(0.7) |
| Theoretical $F=1$ hyperfine shift | $f_{1c,theo}$ | −63,248.0(2.1) |
| Experimental interval $F=0 - F=1$ | $f_{10,exp}$ | 178,254.4(0.9) |
| Theoretical interval $F=0 - F=1$ | $f_{10,theo}$ | 178,247.7(3.3) |
| Theoretical sum of $F=0,1$ hyperfine shifts | $\Sigma_{10}$ | 51,751.7(2.2) |

**References and Notes:**

29. J. C. J. Koelemeij, D. W. E. Noom, D. de Jong, M. A. Haddad, W. Ubachs, Observation of the *v'*=8 ←*v*=0 vibrational overtone in cold trapped HD⁺. *Appl. Phys. B* **107**, 1075–1085 (2012).

30. H. Telle, B. Lipphardt, J. Stenger, Kerr-lens, mode-locked lasers as transfer oscillators for optical frequency measurements. *Appl. Phys. B* **74**, 1-6 (2002).

31. M. Douglas, N. M. Kroll, Quantum electrodynamical corrections to the fine structure of helium. *Ann. Phys.* **82,** 89-155 (1974).

32. K. Pachucki, Quantum electrodynamics effects on helium fine structure. *J. Phys. B: At. Mol. Opt. Phys.* **32,** 137 (1999).

33. M. Haidar, Z.-X. Zhong, V. I. Korobov, J.-Ph. Karr, NRQED approach to the fine- and hyperfine-structure corrections of order $m\alpha^6$ and $m\alpha^6$ (*m/M*): Application to the hydrogen atom. *Phys. Rev. A* **101,** 022501 (2020).

34. D. Bakalov, V. I. Korobov, S. Schiller, Magnetic field effects in the transitions of the HD⁺ molecular ion and precision spectroscopy. *J. Phys. B: At. Mol. Opt. Phys.* **44**, 025003 (2011).

35. S. Schiller, D. Bakalov, A. K. Bekbaev, V. I. Korobov, Static and dynamic polarizability and the Stark and blackbody-radiation frequency shifts of the molecular hydrogen ions H₂⁺, HD⁺, and D2⁺. *Phys. Rev. A* **89**, 052521 (2014).

36. W. Happer, B. S. Mathur, Effective operator formalism in optical pumping. *Phys. Rev.* **163**, 12 (1967).

37. J. C. J. Koelemeij, Infrared dynamic polarizability of HD⁺ rovibrational states. *Phys. Chem. Chem. Phys.* **13,** 18844-18851 (2011).

38. D. Bakalov, S. Schiller, The electric quadrupole moment of molecular hydrogen ions and their potential for a molecular ion clock. *Appl. Phys. B* **114,** 213-230 (2014); **116,** 777-778 (2014).

39. S. Patra, Ph.D. dissertation, Vrije Universiteit Amsterdam (2019).

40. F. Low, Natural line shape. *Phys. Rev.* **88,** 53-57 (1952).

41. T. Udem, L. Maisenbacher, A. Matveev, V. Andreev, A. Grinin, A. Beyer, N. Kolachevsky, R. Pohl, D. C. Yost, T. W. Hänsch, Quantum Interference Line Shifts of Broad Dipole-Allowed Transitions. *Ann. Phys. (Berlin)* **531,** 1900044 (2019).

42. https://physics.nist.gov/cuu/Constants/index.html

43. S. Patra, J.-Ph. Karr, L. Hilico, M. Germann, V. I. Korobov, J. C. J. Koelemeij, Proton–electron mass ratio from HD$^+$ revisited. *J. Phys. B: At. Mol. Opt. Phys.* **51,** 024003 (2018).

44. P. J. Mohr, B. N. Taylor, CODATA recommended values of the fundamental physical constants: 1998. *Rev. Mod. Phys.* **72,** 351-495 (2000).

45. C. G. Parthey *et al.*, Improved Measurement of the Hydrogen 1*S*–2*S* Transition Frequency, *Phys. Rev. Lett.* **107,** 203001 (2011).

46. K. Pachucki, V. Patkóš, V. A. Yerokhin, Three-photon-exchange nuclear structure correction in hydrogenic systems. *Phys, Rev. A* **97,** 062511 (2018).